\newif\ifdraft \draftfalse
\newif\iffull \fullfalse

\newif\ifshort
\newif\ifappendix
\newif\iflabel
\newif\ifcamera

\fulltrue
\camerafalse
\draftfalse

\makeatletter
\@input{tex.flags}
\makeatother

\iffull
\shortfalse
\appendixfalse
\labeltrue
\else
\shorttrue
\appendixfalse
\labelfalse
\fi

\iffull
\documentclass[11pt]{article}
\else
\documentclass{llncs}
\fi

\newcommand{\shortbreak}{{\iffull \else \\ \fi}}

\newcommand{\LABEL}[1]{\iflabel\label{#1}\fi}
\newcommand{\SECTION}[1]{\iflabel\section{#1}\else\section{#1}\fi}
\newcommand{\SUBSECTION}[1]{\iflabel\subsection{#1}\else\subsection{#1}\fi}
\newcommand{\toggleenv}[3]{%
  \newenvironment{#1}[1][]{%
  \iflabel
    \ifx\\##1\\ \begin{#2}\else \begin{#2}[##1]\fi
  \else
    \ifx\\##1\\ \begin{#3}\else \begin{#3}[##1]\fi
  \fi}
  {%
    \iflabel
      \end{#2}
    \else
      \end{#3}
  \fi}
}

\newcommand{\PREF}[1]{\iflabel (\Cref{#1})\xspace \else \unskip\xspace\fi}
\newcommand{\APPREF}[1]{\ifcamera the extended version of the paper\else \Cref{#1}\fi}

\toggleenv{THEOREM}{theorem}{theorem}
\toggleenv{LEMMA}{lemma}{lemma}
\toggleenv{COROLLARY}{corollary}{corollary}
\toggleenv{DEFINITION}{definition}{definition}
\toggleenv{REMARK}{remark}{remark}

\usepackage{algorithm, algorithmic}
\usepackage{amsmath, amssymb}
\usepackage{mathpartir}
\usepackage{moresize}
\usepackage{xspace}
\iffull
\usepackage{fullpage}
\usepackage{amsthm}
\fi
\usepackage{verbatim}
\usepackage{color}
\usepackage{xcolor}
\definecolor{DarkGreen}{rgb}{0.1,0.5,0.1}
\definecolor{DarkRed}{rgb}{0.5,0.1,0.1}
\definecolor{DarkBlue}{rgb}{0.1,0.1,0.5}
\usepackage[numbers]{natbib}
\usepackage{hyperref}
\hypersetup{
    unicode=false,          
    pdftoolbar=true,        
    pdfmenubar=true,        
    pdffitwindow=false,      
    pdftitle={},    
    pdfauthor={}
    pdfsubject={},   
    pdfnewwindow=true,      
    pdfkeywords={keywords}, 
    colorlinks=true,       
    linkcolor=DarkRed,          
    citecolor=DarkGreen,        
    filecolor=DarkRed,      
    urlcolor=DarkBlue,          
}
\usepackage{cleveref}

\newcommand{\jh}[1]{\ifdraft \textcolor{red}{[Justin: #1]}\fi}

\newcommand\R{\mathbb{R}}

\newcommand\cA{\mathcal{A}}

\newcommand\cD{\mathcal{D}}

\newcommand\cK{\mathcal{K}}

\newcommand\cM{\mathcal{M}}

\newcommand\cR{\mathcal{R}}

\newcommand\NN{\mathbb{N}}

\newcommand\RR{\mathbb{R}}

\newcommand{\polylog}{\mathrm{polylog}}

\newcommand\set[1]{\left\{#1\right\}} 

\newcommand{\projs}[1]{\Gamma_{s} #1}
\renewcommand{\tilde}{\widetilde}

\def\<{\langle}
\def\>{\rangle}

\DeclareMathOperator*{\argmin}{argmin}

\DeclareMathOperator*{\OPT}{OPT}

\newcommand{\Lap}{\mathrm{Lap}}
\renewcommand{\bar}{\overline}

\newcommand{\INDSTATE}[1][1]{\STATE\hspace{#1\algorithmicindent}}

\title{Privately Solving Linear Programs\iffull\else\thanks{
  A full version of the paper with the omitted proofs and sections can be found
  at \mbox{\url{http://arxiv.org/abs/1402.3631}}.}\fi
}

\iffull
\author{
  Justin Hsu\thanks{
    Department of Computer and Information Science, University of Pennsylvania.
    Supported in part by NSF Grant CNS-1065060.
  }
  \qquad Aaron Roth\thanks{
    Department of Computer and Information Science, University of Pennsylvania.
    Supported in part by an NSF CAREER award, NSF Grants CCF-1101389 and
    CNS-1065060, and a Google Focused Research Award.
    Email: {\tt aaroth@cis.upenn.edu}.
  }
  \qquad Tim Roughgarden\thanks{
    Department of Computer Science, Stanford University.
    Supported in part by NSF Awards CCF-1016885 and CCF-1215965, and an ONR
    PECASE Award.
  }
  \qquad Jonathan Ullman\thanks{
    School of Engineering and Applied Sciences and Center for Research on
    Computation and Society, Harvard University.  Supported by NSF grant
    CNS-1237235.  Email: {\tt jullman@seas.harvard.edu}.
  }
}
\else
\author{
  Justin Hsu\inst{1}
  \and Aaron Roth\inst{1}
  \and Tim Roughgarden\inst{2}
  \and Jonathan Ullman\inst{3}
  \vspace{-1.0ex}
}

\institute{
  University of Pennsylvania
  \and Stanford University
  \and Harvard University
}
\fi

\begin{document}

\iffull
\newtheorem{theorem}{Theorem}
\newtheorem{lemma}[theorem]{Lemma}
\newtheorem{corollary}[theorem]{Corollary}

\theoremstyle{definition}
\newtheorem{definition}[theorem]{Definition}

\theoremstyle{remark}
\newtheorem{remark}[theorem]{Remark}
\newtheorem{claim}[theorem]{Claim}
\newtheorem{fact}[theorem]{Fact}
\else
\spnewtheorem*{untheorem}{Theorem}{\bfseries}{\itshape}
\spnewtheorem*{unlemma}{Lemma}{\bfseries}{\itshape}
\spnewtheorem*{uncorollary}{Corollary}{\bfseries}{\itshape}
\spnewtheorem*{undefinition}{Definition}{\bfseries}{\itshape}

\spnewtheorem*{unremark}{Remark}{\itshape}{\rmfamily}
\fi

\maketitle
\ifshort\vspace{-3.5ex}\fi
\begin{abstract}
In this paper, we initiate the systematic study of solving linear programs under
differential privacy. The first step is simply
to define the problem: to this end, we introduce several natural classes of
{\em private linear programs} that capture different ways sensitive data can be
incorporated into a linear program. For each class of
linear programs we give an efficient, differentially private solver based on the
multiplicative weights framework, or we give an impossibility result.
\end{abstract}
\iffull
\newpage
\tableofcontents
\fi
\ifshort\vspace{-5ex}\fi

\ifappendix
\else
\SECTION{Introduction} \LABEL{sec:intro}
Linear programming is one of the most fundamental and powerful tools in
algorithmic design.  It is used ubiquitously throughout computer science:
applications include maximum matching, maximum and minimum cost flow, and
fractional packing and covering problems. Linear programming relaxations of
NP-complete problems also underlie countless efficient approximation algorithms.

At the same time, differential privacy is a field where efficient algorithms
have been difficult to find. For many problems in differential privacy, the
initial focus was on understanding the {\em information-theoretic complexity}---the
extent to which solving the problem, efficiently or not, is compatible with
differential privacy.  As a result, there are many central problems that are
known to be privately solvable, but for which computationally efficient
algorithms are not known.  For example, \citet{KLNRS08} show how to privately
PAC learn any PAC learnable concept class (without privacy) with only a small
increase in the sample complexity, but via an exponential time algorithm. It
remains open whether a computationally efficient algorithm can do this in
general.
Similarly, \citet{BLR08} show how to privately release a summary of a private
database that approximately preserves the answers to rich families of linear
queries, again via an exponential time algorithm. In fact, under standard
cryptographic assumptions, it is not possible to efficiently and privately
answer large collections of general linear
queries~\citep{DNRRV09,UV10,Ullman13}.


The two preceding examples are among the many algorithms that use the extremely
general \emph{exponential mechanism} of \citet{MT07} to achieve near optimal
error. However, the exponential mechanism is not efficient in general: it
requires running time linear in the size of its output range, which can be
extremely large. In contrast, general tools for designing \emph{efficient}
differentially private algorithms are harder to come by (although not
non-existent, e.g., the sample and aggregate framework \citep{NRS07} and
output/objective perturbation for unconstrained convex optimization
\citep{CMS11,KST12}).

Our work contributes to the toolbox of general algorithmic techniques for
designing computationally efficient and differentially private algorithms;
specifically, we give tools to privately and efficiently solve linear programs
(LPs) of various types.  An initial problem is to simply \emph{define} what it
means to solve a linear program privately.  Differential privacy is defined in
terms of {\em neighboring databases}.  A database is a collection of records
from some domain and two databases are neighboring if they differ in a single
record.  Differential privacy requires the output distribution of an algorithm
to be nearly identical when run on either of a pair of neighboring databases. If
linear programs can depend on private databases, we naturally have a notion of
{\em neighboring linear programs}, and we want an algorithm for solving these
linear programs that is differentially private with respect to this notion of
neighboring inputs.

The way in which the linear program is derived from the database gives rise to
several distinct notions of neighboring linear programs. For instance, consider
an LP with objective $c^\top x$ and constraints $Ax \leq b$, where moving to a
neighboring LP neighboring database leaves $c$ and $A$ unchanged but perturbs
$b$ by only a small amount in each coordinate.  Solving this kind of linear
programming privately is similar to the well-studied {\em linear query release}
problem in differential privacy, and techniques for linear query release---such
as the private multiplicative weights algorithm of \citet{HR10} (and its offline
variants \citep{GRU12,HLM12})---can be adapted with minor changes. (This result
may even be considered folklore.)  On the other hand, the situation is
qualitatively different if moving to a neighboring LP can change either the
constraint matrix $A$ or the objective vector $c$. Some of these private LPs can
still be solved; others are provably impossible to solve to nontrivial accuracy
under differential privacy.

In this paper, we develop a taxonomy of private LPs.  For each class, we either
present an efficient and accurate differentially private solver, or prove that
general LPs of this type cannot be accurately solved while preserving privacy.

\SUBSECTION{Our Results and Techniques}
We consider linear programs $LP(D)$ defined by a database $D$, with form
\begin{align*}
  &\max_{x \in \RR_+^d} c^\top x \\
  \text{s.t. } &Ax \leq b.
\end{align*}

Here, the vector $x$ represents the variables of the linear program, and $c =
c(D), A = A(D)$, and $b = b(D)$ may each depend on the private database $D$. Our
goal is to find an approximate solution to $LP(D)$ in a sense to be defined,
while ensuring differential privacy for the underlying database $D$.

We classify private LPs along two dimensions: {\em which part} of the LP depends
on the database and {\em how sensitive} the LP is to changes in the database.
Along the second axis, we will consider: 1) \emph{low-sensitivity} LPs, where
changing one record of the database induces a small difference between
coefficients that vanishes as the size of the database $n$ grows and 2)
\emph{high-sensitivity} LPs, where changing one record of the database can
induce a potentially large change in some coefficient.  Low-sensitivity LPs are
natural when the coefficients of the LP represent some kind of average over the
database, whereas high-sensitivity LPs are natural when the coefficients
represent specific records of the database.

Furthermore, we consider four parts of the LP that might depend on the database:
1) \emph{the rows of $A$}, 2) \emph{the scalars $b$}, 3) \emph{the columns of
  $A$}, and 4) \emph{the objective $c$}.  These four parts of the LP, combined
with the two notions of sensitivity, lead to the following eight notions of
private linear programming:


\begin{enumerate}
  \item \textbf{The constraints:} For these linear programs, moving to a
    neighboring \shortbreak database can affect at most one row of $A$ and the
    corresponding entry of $b$, which corresponds to changing one constraint of
    the LP.

    \begin{enumerate}
      \item \textbf{High-sensitivity:} For \emph{high-sensitivity constraint
          private LPs} \PREF{sec:cp}, moving to a neighboring database can
        change a single constraint arbitrarily.  That is, for every pair of
        neighboring databases $D, D'$, there exists a row $i$ such that for
        every row $j \neq i$, $A(D)_j = A(D')_j$ and $b(D)_j = b(D')_j$.  This
        kind of linear program arises, for example, in covering LPs in which
        each record of the database represents an individual that needs to be
        covered.  We cannot hope to approximately satisfy \emph{every}
        constraint while ensuring privacy,%
        \ifshort\else\footnote{%
          For example, given a solution $x$ to $LP(D)$, we can always derive a
          new constraint $(A_i, b_i)$ that is far from being satisfied by $x$.
          If we introduce this new constraint in a neighboring linear program
          $LP(D')$, by the differential privacy condition, this new constraint
          must also be far from being satisfied in $LP(D')$ with high
          probability.}\fi{}
        but we show that by using a variant of multiplicative weights that
        operates only over a restricted set of distributions, we can still find
        solutions to such LPs that approximately satisfy \emph{most} of the
        constraints. As an example of our technique, we solve a private version
        of the fractional set cover problem.
      \item \textbf{Low-sensitivity:} For \emph{low-sensitivity constraint
          private LPs} \PREF{sec:mp}, moving to a neighboring database can
        change a single row of $A$ by a small amount in each entry---for some
        row $i$, $\|A_i(D) - A_i(D') \|_\infty \leq 1/n$. We show how to solve
        these LPs using multiplicative weights; our techniques work equally well
        if the entire constraint matrix can change on neighboring problems (we
        will sometimes call these {\em low-sensitivity matrix} or {\em row
          private LPs}).
    \end{enumerate}

  \item \textbf{The scalars:} For these linear programs, $c$ and $A$ are fixed
    and moving to a neighboring database only affects $b = b(D)$.
  
    \begin{enumerate}
      \item \textbf{High-sensitivity:} For \emph{high-sensitivity scalar private
          LPs}, for every neighboring $D,D'$, there is a row $i$ such that for
        every $j \neq i$, $b(D)_j = b(D')_j$.  We show that in general, such LPs
        cannot be solved privately.
      \item \textbf{Low-sensitivity:} For \emph{low-sensitivity scalar private
          LPs}, moving to a neighboring database can change every entry in $b$
        slightly, such that $\|b(D) - b(D')\|_{\infty} \leq 1/n$.  These LPs
        capture the \emph{private linear query release} problem, so we will
        sometimes refer to them as \emph{query release LPs}.    In this problem,
        the database is viewed as a histogram $D \in \NN_+^d$ and the objective
        is to find a {\em synthetic database} $x \in \R_+^d$ such that for every
        linear query $q$ in some family, $\< q,  x \> \approx \<q,  D\>$. We
        show how to adapt existing techniques for this problem and derive
        resulting accurate solvers for LPs of this form \PREF{sec:qr}.
    \end{enumerate}
  \item \textbf{One column in $A$}: For these linear programs, moving to a
    neighboring database can affect at most one column of $A$.
    
    \iffull
    These LPs arise literally as the dual linear programs of row private LPs.
    For example, in a LP where variables represent different tasks, the private
    coefficients corresponding to a single variable may represent the amount of
    resources needed for that task.  Then, a packing LP seeks to maximize some
    objective subject to resource constraints.
    \fi

    \begin{enumerate}
      \item {\bf High-sensitivity:} For {\em high-sensitivity column private
          LPs}, for every neighboring $D, D'$, the matrices $A(D), A(D')$ are
        arbitrarily different in a single column, and identical in all other
        columns. We show that in general, such LPs cannot be solved privately
        \PREF{sec:highsens-constr}.
      \item {\bf Low-sensitivity:} For {\em low-sensitivity column private LPs},
        moving to a neighboring database can change every entry in a single
        column of $A$ by a small amount. More generally, if $A_i$ is the $i$th
        {\em row} of $A$, then $\|A(D)_i - A(D')_i \|_1 \leq 1/n$ for each $i$. We
        show how to use these LPs using multiplicative weights \PREF{sec:colp}.
    \end{enumerate}

  \item \textbf{The objective}: For these linear programs, moving to a
    neighboring database can affect the objective $c$. The scalars $b$ and
    constraints $A$ remain unchanged.
    
    \begin{enumerate}
      \item {\bf High-sensitivity:} For {\em high-sensitivity objective private
          LPs}, for every neighboring $D, D'$, a single entry of the objective
        $c(D), c(D')$ can change arbitrarily. We show that in general, such LPs
        cannot be solved privately \PREF{sec:highsens-obj}.

      \item {\bf Low-sensitivity:} For {\em low-sensitivity objective private
          LPs}, for every neighboring $D, D'$, the objective vectors $c(D),
        c(D')$ satisfy $\|c(D) - c(D')\|_1 \leq 1/n$.  This kind of linear
        program can be solved inefficiently to high accuracy by selecting from
        the set of vertices of the feasible polytope with the exponential
        mechanism; we show that linear programs in this class can also be solved
        efficiently and accurately, by directly using randomized response
        \PREF{sec:op}.
    \end{enumerate}
\end{enumerate}

This taxonomy is summarized in \Cref{tbl:big}. We will formally define accuracy,
but roughly speaking, an accurate solution satisfies each constraint to within
additive $\alpha$, and has objective within additive $\alpha$ of optimal (when
there is an objective). The exception is constraint privacy (indicated by the
asterisk), where our algorithm finds a solution that satisfies only {\em most}
of the constraints to within additive $\alpha$, and may violate the other
constraints arbitrarily.
\begin{table}[ht!]
  \centering
  \begin{tabular}{|l|l|l|}
    \hline
    Location of change     & High sensitivity          & Low sensitivity       \\
    \hline
    \hline
    Objective $c$          & No \PREF{sec:lowerbounds} & Yes \PREF{sec:op}   \\
    \hline
    Scalar $b$             & No \PREF{sec:lowerbounds} & Yes \iflabel(Folklore,
    \Cref{sec:qr})\fi   \\
    \hline
    Row/All of $A$         & Yes* \PREF{sec:cp}        & Yes \PREF{sec:colp} \\
    \hline
    Column of $A$          & No \PREF{sec:lowerbounds} & Yes \PREF{sec:colp} \\
    \hline
  \end{tabular}
  \caption{Efficient and accurate solvability}
  \label{tbl:big}
\end{table}
%
%
\ifshort\vspace{-10ex}\fi
\SUBSECTION{Related Work} \LABEL{sec:related}
Differential privacy emerged from a line of work initiated by \citet{DN03}, was
defined by \citet{DMNS06}, and is now a standard definition of privacy in
computer science.  Below, we discuss relevant results in differential privacy;
the survey by~\citet{DworkSurvey} is an excellent source for a more
comprehensive overview.

Private optimization has been studied since the work of \citet{BDMN05} and
\citet{KLNRS08}, who considered how to choose an optimal classifier privately.
\citet{BDMN05} give an efficient reduction from SQ learning to private SQ
learning, and \citet{KLNRS08} give a very general but inefficient reduction from
PAC learning to private PAC learning using the exponential mechanism of
\citet{MT07}.  Private learning was placed explicitly into an optimization
framework by \citet{CMS11}, who give two techniques for privately solving
certain \emph{unconstrained} convex optimization problems.  \citet{GLMRT10} give
several algorithms for problems in private \emph{combinatorial optimization},
but these were specialized combinatorial algorithms for specific problems.

In parallel, a line of work initiated by \citet{BLR08} and continuing with
\citet{DNRRV09,RR10,DRV10,HR10,GRU12,HLM12} study the problem of privately
producing \emph{synthetic data} consistent with some private database on many
\emph{linear queries}. (Of particular note is the private multiplicative weights
mechanism of \citet{HR10}, which achieves the optimal accuracy and running
time bounds~\citep{Ullman13,BUV14}.) This problem can be represented as a linear
program with queries defining constraints, and indeed, the private
multiplicative weights algorithm of \citet{HR10} can be directly applied to
solve this kind of linear program. This observation motivates our current
investigation.

Our algorithms are mostly based on different variants of the {\em multiplicative
  weights} method of solving linear programs, which was introduced by
\citet{PST95} (see the excellent survey by \citet{AHK12} for more details).
Whereas \citet{PST95} maintain a distribution over the dual variables with
multiplicative weights, depending on the kind of linear program we are solving,
we either maintain a distribution over the dual variables or the primal
variables. To solve \emph{constraint private} LPs, we use a combination of the
multiplicative weights update method and {\em Bregman projections}
\citep{AHK12}---\citet{HRU13} use a similar version of this technique in
designing {\em analyst private} mechanisms.


\fi

\SECTION{Differential Privacy Preliminaries} \LABEL{sec:prelim}

\ifshort
We defer differential privacy preliminaries to \APPREF{sec:prelim}.
\else
Differential privacy is a strong notion of privacy, first introduced
by~\citet{DMNS06}. In the typical setting, we consider a {\em database} as a
multisets of records, each belonging to a single individual. Then, a randomized
function from databases to an output range satisfies differential privacy if,
for any change in a single record of the input database, the distribution on
outputs remains roughly the same. More formally, we have the following
definition.

\begin{DEFINITION}[\citet{DMNS06}]
  Let $\epsilon > 0$ and $0 \leq \delta < 1$ be given. A randomized function $M
  : \cD \rightarrow \cR$ mapping databases to an output range is {\em
  $(\epsilon, \delta)$-differentially private} if for every subset $S \subseteq
  \cR$ and for every pair of database $D, D'$ that differ in a single record,
  \[
    \Pr[ M(D) \in S ] \leq e^\epsilon \Pr[ M(D') \in S ] + \delta.
  \]
  When $\delta = 0$, we will say that $M$ is {\em $\epsilon$-differentially
    private}.
\end{DEFINITION}

We will use two basic mechanisms from differential privacy: the Laplace
mechanism and the exponential mechanism. The Laplace mechanism privately
releases a number by adding noise drawn from the Laplace distribution.

\begin{DEFINITION}[\citet{DMNS06}]
  Let $\epsilon > 0$ be given. A function $f : \cD \rightarrow \RR$ is {\em
  $\Delta$-sensitive} if for every pair of database $D, D'$ that differ in a
single record,
  \[
    |f(D) - f(D')| \leq \Delta.
  \]
  The {\em Laplace mechanism} applied to a $\Delta$-sensitive function releases
  \[
    f(D) + \nu,
  \]
  where $\nu$ is a draw from the {\em Laplace distribution} with parameter
  $\epsilon/\Delta$; that is, with probability density function
  \[
    F(\nu) = \frac{\epsilon}{2\Delta} \exp \left( - \frac{|\nu|
        \epsilon}{\Delta} \right).
  \]
\end{DEFINITION}

The Laplace mechanism is $\epsilon$-differentially private and satisfies the
following tail bound, which is also an accuracy guarantee for the Laplace
mechanism.

\begin{LEMMA} \LABEL{lap:tail}
  Let $\beta \in (0, 1)$ be given, and let $\nu$ be drawn from the Laplace
  distribution with scale $b$. Then,
  \[
    \Pr[ |\nu| \geq T ] \leq \beta
    \quad \text{for} \quad
    T = \frac{1}{b} \log(1/\beta).
  \]
\end{LEMMA}

We will also use the exponential mechanism \citep{MT07}, which can privately
produce a non-numeric or discrete output.  The exponential mechanism is defined
in terms of a \emph{quality score} that maps a database and an element of the
range to a real valued score.  For a given a database, the exponential mechanism
privately outputs an element of the range that approximately maximizes the
quality score.

\begin{DEFINITION}[\citet{MT07}]
  Let $\epsilon > 0$ be given, and suppose the {\em quality score} $Q : \cR
  \times \cD \rightarrow \RR$ is $\Delta$-sensitive in the database.
  On database $D$, the {\em $\epsilon$-private exponential mechanism with
  quality score $Q$} outputs $r \in \cR$ with probability proportional to
  \[
    \exp \left( \frac{\epsilon}{2 \Delta} \cdot Q(r, D) \right).
  \]
\end{DEFINITION}

The exponential mechanism is $\epsilon$-differentially private, and satisfies
the following accuracy guarantee.

\begin{THEOREM}[\citet{MT07}] \LABEL{exp:acc}
  Let $\beta \in (0, 1)$ and the database $D$ be given. Suppose that the maximum
  value of the quality score $Q$ on database $D$ is $\OPT$. Then, the
  $\epsilon$-private exponential mechanism with quality score $Q$ on $D$ outputs
  $r \in \cR$ such that
  \[
    \Pr\left[ Q(r, D) \geq \OPT - \frac{2 \Delta}{\epsilon} \log \left(
        \frac{|\cR|}{\beta} \right) \right] \geq 1 - \beta.
  \]
\end{THEOREM}

To combine these mechanisms, we will use standard composition theorems.

\begin{THEOREM}[\citet{DRV10}] \LABEL{composition}
  For any $\delta \in (0, 1)$, the composition of $k$ (adaptively chosen) $\epsilon'$-private
  mechanisms is $(\epsilon, \delta)$-differentially private, for
  \[
    \epsilon' = \frac{\epsilon}{\sqrt{8k \log (1/\delta)}}.
  \]
\end{THEOREM}
\fi

\SECTION{Constraint Private LPs} \LABEL{sec:cp}
\ifappendix
\else
Let us begin by considering {\em constraint private LPs}, with the general form
\begin{align*}
  &\max_{x \in \cK} c^\top x \\
  \text{s.t. } &Ax \leq b,
\end{align*}
where $A \in \RR^{m \times d}, b \in \RR^m, c \in \RR^d$, and $\cK \subseteq
\RR^d$.  We think of $\cK$ as the {\em easy} constraints, those that are
independent of the database, like non-negativity.

Let $\cK_{\OPT} = \cK \cap \{ x \in \RR^d \mid c^\top x = \OPT \}$. Then, the
original LP can be solved approximately by repeatedly solving the feasibility
problem
\begin{align*}
  &\text{find } x  \in \cK_{\OPT} \\
  \text{s.t. } &Ax \leq b,
\end{align*}
binary searching on the optimal objective value $\OPT$.%
\footnote{%
  Binary search will incur an additional overhead in privacy, but in some
  situations may not be necessary: for instance, if a bound on the sensitivity
  of the optimal objective is known, we can solve the LP non-privately and
  estimate $\OPT$ with the Laplace mechanism.}{}
Thus, unless we specify otherwise, we will restrict our attention to feasibility
LPs. Furthermore, since a linear program has a convex feasible region,
$\cK$ (and hence $\cK_{\OPT}$) are convex. From now on, we will write $\cK$ for
$\cK_{\OPT}$.
\fi

\ifshort
For constraint privacy, we want to find a solution that hides whether a single
constraint is in the LP or not.
\else
Roughly, a private database $D$ defines a linear program with objective vector
$c(D)$, constraint matrix $A(D)$, and vector $b(D)$, which we will call the {\em
  scalars}. In a constraint private LP, the objective $c(D) = c(D')$ is
independent of the data and for every two neighboring datasets $D, D'$,  the
matrices $A(D)$ and $A(D')$ are exactly the same, except one matrix has an
additional row that the other does not. The vectors $b(D)$ and $b(D')$ are also
identical, except there is an  entry corresponding to the different constraint
in $A$ in one $b$, but not the other.  Then, we want a LP solver that satisfies
the differential privacy guarantee with respect to this notion of adjacency.
\fi%
Formally:

\begin{DEFINITION}
  A randomized algorithm $\cM$ with inputs $m \in \NN$, vector $b \in \RR^m$,
  and matrix $A \in \RR^{m \times d}$ and outputting a vector in $\RR^d$ is {\em
    $(\epsilon, \delta)$-high sensitivity constraint private} if for any $A, A'$
  such that $A'$ is equal to $A$ with an additional row appended, and $b, b'$
  such that $b'$ is equal to $b$ with an additional entry,
  \[
    \Pr[ \cM(m, b, A) \in S ] \leq e^\epsilon \Pr[ \cM(m + 1, b', A') \in S] +
    \delta 
  \]
  for any set $S \subseteq \RR^d$.
\end{DEFINITION}

\SUBSECTION{Solving LPs with Dense Multiplicative Weights}
A standard approach to solving LPs is via {\em no-regret} algorithms.
\ifshort
\else
For a brief summary, these algorithms operate over a series of timesteps,
selecting a single action at each step. Once the action is selected, the loss
for each action is revealed (perhaps adversarially); the no-regret algorithm
then adjusts to favor actions with less loss.
\fi{}
While LPs can be solved using any no-regret algorithm, for concreteness we use
the multiplicative weights update algorithm.

\ifshort
\else
Throughout, we will use calligraphic letters ($\cA$) to denote sets of actions,
Roman letters $(A)$ to denote measures on those actions $A : \cA \rightarrow [0,
1]$, and letters with tildes ($\tilde{A}$) to denote a probability
distributions over actions. We will write $|A|$ to mean the {\em density} of
measure $A$, defined to be $\sum_{a \in \cA} A_a$.
\fi

We will use a variant of the standard multiplicative weights algorithm that
maintains a {\em dense distribution} over the set of constraints, i.e., a
distribution that doesn't place too much probability on any action.  We will
call this algorithm, due to~\citet{HW01}, the {\em dense multiplicative weights
  algorithm} (\Cref{alg:DMW}). Roughly, the algorithm projects the MW
distribution on actions into the set of dense distributions at each step. The
loss at each step will be defined by a point that approximately satisfies the
average constraint weighted by the MW distribution---by capping the probability
on any constraint, we ensure that this point can be selected privately even when
a single constraint can change arbitrarily on neighboring instances.

We first define this projection step, also known as a {\em Bregman projection}.

\begin{DEFINITION} \LABEL{def:proj}
  Let $s > 0$. Given a measure $A$ such that $|A| \leq s$, let $\projs{A}$ be
  the {\em (Bregman) projection of $A$ into the set of $1/s$-dense
  distributions}, defined by $\projs{A}_a = \frac{1}{s} \cdot \min\{1, c A_a\}$
  for every $a \in \cA$, where $c \geq 0$ is such that $s = \sum_{a \in \cA}
  \min\{1, c A_a\}$.
\end{DEFINITION}

Then, we can define the Dense Multiplicative Weights algorithm, which uses the
standard multiplicative weights update rule combined with a Bregman projection
into the set of dense distributions after each step.

\begin{algorithm}[ht!]
  \begin{algorithmic}
    \STATE{Let $A_1$ be the uniform measure on $\cA$}
    \STATE{For $t = 1,2,\dots,T$:}
    \INDSTATE[1]{Let $\tilde{B}^{t} = \projs{A^t}$}
    \INDSTATE[1]{Receive loss vector $\ell^t$ (may depend on $B^{1}, \dots,
    B^{t}$)}
    \INDSTATE[1]{\textbf{Update:} {\bf For each} $a \in \cA${\bf :}}
    \INDSTATE[2]{Update $A^{t+1}_a = e^{- \eta \ell^t_a} A^{t}_a$}
  \end{algorithmic}
  \caption{The Dense Multiplicative Weights algorithm, $DMW_{s, \eta}$}
  \label{alg:DMW}
\end{algorithm}

\ifshort
\else
Then, dense multiplicative weights satisfies the following (regret)
guarantee.%
\footnote{%
  Note that the regret guarantee is only with respect to dense distributions,
  rather than arbitrary distributions. This is a result of projecting the MW
  distribution to a dense distribution---the algorithm may not be able to
  compete with non-dense distributions.}

\begin{THEOREM}[\citet{HW01}] \LABEL{DMW-regret}
  Let $A_1$ be the uniform measure of density $1$ and let $\set{\tilde{B}^t}$
  be the sequence of projected distributions obtained by $DMW_{s,\eta}$ with
  arbitrary losses $\set{\ell^t}$ satisfying $\| \ell^t \|_\infty \leq 1$ and
  $\eta \leq 1/2$.  Let $\tilde{B}^*$ be the uniform distribution on some subset
  $S^* \subseteq \cA$ of size $s$.  Then,
  \begin{align*}
    \frac{1}{T} \sum_{i = 1}^T \< \ell^t, \tilde{B}^t \>
    &\leq
    \frac{1}{T} \sum_{i = 1}^T \< \ell^t, \tilde{B}^* \>  + \eta + \frac{\log
      |\cA|}{\eta T}.
  \end{align*}
\end{THEOREM}
\fi
Recall we can assume that we know the optimal value $\OPT$, so the objective can
be represented as the constraint $c^\top x = \OPT$. Hence, let $\cK = \{ x \in
\RR^d_+ \mid c^\top x = \OPT \}$ be the public feasible set.  We will assume
that there is a known, data-independent upper bound $\rho$ such that
\[
  \rho \geq \max_D \max_{x \in \cK} \| A(D)x - b(D) \|_\infty,
\]
which we call the {\em width} of the LP.

We will define our algorithm in terms of an approximate oracle for solving a
linear minization problem. (For a concrete example of such an oracle in the
context of fractional set cover, see the next section.)
\begin{DEFINITION}
  An {\em $(\alpha, \beta)$-approximate, $\rho$-bounded oracle}, given a
  distribution $y \in \RR^m$ and matrix $A \in \RR^{m \times d}$, with
  probability at least $1-\beta$ finds $x^* \in \RR^d$ with
  \[
    \sum_{i = 1}^{m} y_i (A_i \cdot x^*) \leq  \min_{x \in \cK} \sum_{i = 1}^{m}
    y_i (A_i \cdot x) + \alpha
  \]
  and $\|A x^* - b\|_\infty \leq \rho$.
\end{DEFINITION}

\ifshort
We present the full algorithm in \Cref{dual-lp-nonpriv}.
\else
To solve linear programs, we use the dense multiplicative weights algorithm to
maintain a distribution over the constraints, and pick points $x^t \in \cK$ that best
satisfy the weighted combination of constraints at each step. Intuitively, the
losses will lead to more weight on violated constraints, leading to points that
are more feasible. Taking the average of the points $x^t$ will yield an
approximately feasible point, if it exists.  See \Cref{dual-lp-nonpriv} for the
full algorithm.

We note that similar techniques for solving linear programs using multiplicative
weights have been known since at least \citet{PST95}; the novelty in our
approach is that we use multiplicative weights paired with a projection onto the
set of dense distributions, and show that the solution approximately satisfies
\emph{most} of the constraints. As we will see, the projection step is needed
for privacy.
\fi

\begin{algorithm}[ht!]
  \begin{algorithmic}
    \STATE{Input $A \in \RR^{m \times d}$, $b \in \RR^m$.}
    \STATE{Let $\tilde{y}^{1}$ be the uniform distribution in $\RR^m$, $\rho
      \geq \max_{x \in \cK} \|Ax - b\|_\infty$ be the {\em width} of the LP,
      $s \in \NN$ be the {\em density parameter}, and $\alpha > 0$ be the
      desired accuracy. Let $\mathit{Oracle}$ be an $(\alpha, \beta)$-accurate,
      $\rho$-bounded oracle, and set
      \begin{mathpar}
        \eta = \sqrt{\frac{\log m}{T}}, \and
        T = \frac{36 \rho^2 \log m}{\alpha^2}.
      \end{mathpar}
    }
    \STATE{For $t = 1,\dots,T$:}
    \INDSTATE[1]{Find $x^{t} = \mathit{Oracle}(\tilde{y}^t, A)$}
    \INDSTATE[1]{Compute losses $\ell^t_i := (1/2\rho) (b_i - A_i \cdot x^t) +
      1/2$. }
    \INDSTATE[1]{Update $\tilde{y}^{t+1}$ from $\tilde{y}^t$ and $\ell^t$ via
      dense multiplicative weights with density $s$.}
    \STATE{Output $\bar{x} = (1/T) \sum_{t = 1}^T x^t$.}
  \end{algorithmic}
  \caption{Solving for LP feasibility with dense multiplicative weights}
  \label{dual-lp-nonpriv}
\end{algorithm}

\ifshort
If the oracle is sufficiently accurate, the following theorem bounds the number
of iterations needed to achieve a set level of accuracy.
\fi
\begin{theorem} \label{dual-lp}
  Let $0 < \alpha \leq 9 \rho$, and let $\beta \in (0, 1)$.  Suppose there is
  a feasible solution of the linear program.  Then with
  probability at least $1 - \beta$, \Cref{dual-lp-nonpriv} with density
  parameter $s$ run with an $(\alpha/3, \beta/T)$-approximate, $\rho$-bounded
  oracle finds a point $x^*$ in $\cK$ such that there is a set of
  constraints $S$ of size at most $|S| < s$, with $A_i x^* \leq b_i + \alpha$
  for every $i \notin S$.
\end{theorem}
\ifshort
\else
\begin{proof}
  By a union bound over $T$ steps, the oracle succeeds on all steps with
  probability at least $1 - \beta$; condition on this event.

  Let $\cK_s  = \{ y \in \R^m \mid \mathbf{1}^\top y, \|y\|_\infty \leq 1/s \}$
  be the set of $1/s$-dense distributions. Then,  $y^\top A x^* \leq y^\top
  b$ for any $y \in \cK_s$, so in particular the oracle finds $x^t$ with
  $y^\top A x^t < y^\top b + \alpha/3$.

  Thus, the loss vectors $\ell^t = (1/2\rho) (b - Ax^t) + 1/2$ satisfy $\ell^t
  \cdot y^t \geq 1/2 - \alpha/6  \rho$, which is at least $-1$ if $\alpha \leq
  9  \rho$. Since the oracle is $\rho$-bounded, $\ell^t \cdot y^t \leq 1$. So,
  \Cref{DMW-regret} applies; for $p$ any point in $\cK_s$, we have the following
  bound:
  \begin{align*}
    \frac{1}{2} - \frac{\alpha}{6\rho} &\leq \frac{1}{T}\sum_{t = 1}^T \left(
    \ell^t \cdot p \right) + \eta + \frac{\log m}{\eta T} \\
    &= \frac{1}{T}\sum_{t = 1}^T \left( \frac{1}{2\rho} \left( b - A
        x^t \right) + \frac{1}{2} \right)\cdot p + \eta + \frac{\log m}{\eta T}
  \end{align*}
  Thus,
  \[
    -\frac{\alpha}{6\rho}  \leq \frac{1}{T} \sum_{t = 1}^T \frac{1}{2\rho}
    \left( b - A x^t \right) \cdot p + \eta + \frac{\log m }{\eta T}.
  \]
  Define $x = (1/T) \sum_{t = 1}^T x^t$, and rearrange:
  \[
    p^\top  Ax \leq p^\top b + 2 \rho \eta + \frac{2  \rho \log m }{\eta T} +
    \frac{\alpha}{3}.
  \]
  By our choice of $\eta$ and $T$, we get
  \[
    p^\top Ax \leq p^\top b + \alpha.
  \]
  Since this holds for any $p \in \cK_s$, $x$ satisfies all but $s-1$
  constraints with error $\alpha$---if it didn't, letting $p$ be the uniform
  distribution on the $s$ violated constraints would give a contradiction.
\end{proof}

\SUBSECTION{Achieving Constraint Privacy}
Now, we will see how to make \Cref{dual-lp-nonpriv} constraint private. First,
the output point depends on the private data (the constraints $A$) only through
the minimization step. Thus, if we can make the minimization private (in a
certain sense), then each $x^t$ (and hence the final point $\bar{x}$) will
satisfy constraint privacy. Note that if the oracle privately minimizes over
$\cK$, the final point $\bar{x}$ will automatically be in $\cK$ since $\cK$ is
convex. Hence, we can also think of $\cK$ as the {\em public} constraints, the
ones that are always satisfied.
\fi
\ifshort
We also have the following privacy guarantee.
\fi
\begin{THEOREM}\LABEL{cp-priv}
  Let $\epsilon, \delta, T > 0$, and let
  \[
    \epsilon' = \frac{\epsilon}{\sqrt{8T \log(1/\delta)}}.
  \]
  with density parameter $s \in \NN$. Suppose the oracle is
  $\epsilon'$-private, where on neighboring instances the inputs (distributions)
  $\tilde{y}, \tilde{y}'$ satisfy
  \begin{mathpar}
    \|\tilde{y}\|_\infty \leq 1/s, \and
    \|\tilde{y}'\|_\infty \leq 1/s, \and
    \|\tilde{y} - \tilde{y}'\|_1 \leq 2/s,
  \end{mathpar}
  and the matrices $A, A'$ are exactly the same except one has an additional
  row, and the vectors $b, b'$ except one has a corresponding additional entry.
  Then, \Cref{dual-lp-nonpriv} with density $s$ is $(\epsilon, \delta)$-high
  sensitivity constraint private.
\end{THEOREM}
\ifshort
\else
\begin{proof}
  If the oracle is $\epsilon'$-differentially private, then $(\epsilon,
  \delta)$-constraint privacy for the whole algorithm follows directly by
  composition (\Cref{composition}).

  To show that the oracle is private when adding or removing a constraint from
  the LP, we know that $A, A'$ are exactly the same except one has an extra row,
  and we know that $\|\tilde{y}\|_\infty \leq 1/s$ since we have projected into
  the set $\cK_s$.  Hence, it only remains to check that neighboring $\tilde{y},
  \tilde{y}'$ satisfy $\| \tilde{y} - \tilde{y}' \|_1 \leq 2/s$ for each
  timestep $t$. We use a result about the sensitivity of Bregman projections
  from from~\citet{HRU13}; we reproduce the proof for completeness.

  \begin{LEMMA}[\citet{HRU13}]
    Let $s > 0$ be given.  Suppose $A, A'$ be measures on sets $\cA, \cA \cup a'$
    respectively, and identical on $\cA$. If $\tilde{A}, \tilde{A}'$ are the
    respective Bregman projections into the set of $1/s$-dense distributions,
    then
    \[
      \| \tilde{A} - \tilde{A}' \|_1 \leq 2/s.
    \]
    Here and below, we treat $A, A'$ as supported on the same set with $A_{a'}$
    fixed at $0$.
  \end{LEMMA}
  \begin{proof}
    From the definition of the projection (\Cref{def:proj}), it's clear that $s
    \tilde{A}_a \geq s\tilde{A}'_a$ for all $a \neq a'$. We then have the
    following:
    \begin{align*}
      \sum_{a \in \cA \cup a'} |s \tilde{A}_a - s \tilde{A}'_a|
      &= |s\tilde{A}_{a'} - s \tilde{A}'_{a'}| + \sum_{a \neq a'} |s\tilde{A}_a -
      s\tilde{A}'_a| \\
      &\leq 1 + \sum_{a \neq a'} |s\tilde{A}_a - s\tilde{A}'_a| \\
      &= 1 + \sum_{a \neq a'} s\tilde{A}_a - s\tilde{A}'_a \\
      &= 1 + s|\tilde{A}'| - s(|\tilde{A}'| - \tilde{A}'_{a'}) \\
      &\leq 1 + s - (s - 1) = 2
    \end{align*}
    Dividing through by $s$, we are done.
  \end{proof}

  Since $y, y'$ are identical except for the weight corresponding
  to the differing constraint, we are done by the lemma.

\end{proof}
\fi

Now that we have presented our algorithm for solving LPs under constraint
privacy, we give an example of how to instantiate the oracle and apply
\Cref{dual-lp}.

\SUBSECTION{Private Fractional Set Cover}
\ifappendix
\else
We will consider the example of the {\em fractional set cover} LP, though our
arguments extend to constraint private LPs with a private oracle that has low
width. (For example, many covering and packing LPs satisfy this property.)

Suppose there are $d$ sets, each covering some subset of $m$ people. Each set
has a cost $c_S$, and we wish to select the cheapest collection of sets that
covers every person. We will consider the fractional relaxation of this problem,
where instead of selecting whole sets for the cover, we can decide to select a
fraction of each set, i.e., each set can be chosen to some non-negative degree,
and the cost for set $S$ is the degree to which it is open times $c_S$.
We again want the cheapest fractional collection of sets, such that at least
weight $1$ covers each person.%
\footnote{%
  To highlight the constraint private LP, we will only consider the fractional
  version. It is also possible to round the fractional solution to an integral
  solution (with slightly worse cost), since randomized rounding is independent
  of the private data. }

To formulate this as a linear program, let the variables be $x \in \RR^d_+$;
variable $x_S$ will be the degree that we choose set $S$ in the cover. For the
constraints, let $A_i \in \{0, 1\}^m$ such that $A_{iS}$ is $1$ exactly when set
$S$ covers $i$, otherwise $0$.

We will assume that the optimal value $\OPT$ is known, and the goal is to compute
an approximate fractional set covering $x^*$ corresponding to $\OPT$. This is
equivalent to solving the following linear program:
\begin{align*}
  &\text{find: } x \in \cK \\
  \text{s.t. } &A_i \cdot x \geq 1 \quad \text{for each } i \\
\end{align*}
where $\cK = \{ x \in \RR^d_+ \mid c \cdot x = \OPT \}$ is the feasible region.


We wish to achieve constraint privacy: if each individual corresponds to a
covering constraint, then we want an approximate solution that is hides whether
a person $i$ needs to be covered or not. This is not always possible---if each
set contains just one person, then the presence of a set in any valid covering
will reveal information about the people that need to be covered.  Thus, we will
find a solution violating a few constraints, so only covering {\em most} people.

To use our constraint private LP solver, we first define a private oracle
solving the minimization problem
\[
  O(y) = \argmin_{x \in \cK} \sum_i y_i (A_i \cdot x).
\]
Since the oracle is minimizing a linear function, the optimal point lies at a
vertex of $\cK$ and is of the form
\[
  x^* = \frac{\OPT}{c_i} e_i
\]
for some $i$, where $e_i$ is the $i$'th standard basis vector, i.e., all zeros
except for a $1$ in the $i$'th coordinate. We can use the exponential mechanism
to privately select this vertex.
\fi
\ifappendix
We will show that the exponential mechanism is a suitable oracle.
\fi
\ifshort
\else
\begin{LEMMA} \LABEL{sc-oracle}
  Let $\gamma \in (0, 1)$ be given. Suppose $\|y\|_\infty \leq 1/s$, and
  suppose that $\|y - y'\|_\infty \leq 2/s$ on adjacent inputs. Let $O(y)$
  be the $\epsilon$-private exponential mechanism over the vertices of $\cK$
  with quality score
  \[
    Q(j, y) = \sum_i y_i \left(A_i \cdot \frac{\OPT}{c_j} e_j  - 1 \right) =
    \frac{\OPT}{c_j} \sum_i y_i a_{ij} - 1.
  \]
  Then $O$ is an $(\alpha, \gamma)$-approximate, $\rho$-bounded oracle,
  with
  \[
    \rho = \frac{\OPT}{c_{min}} - 1
    \quad \text{and} \quad
    \alpha = \frac{ 6 \OPT \log d \log(1/\gamma)}{c_{min} \cdot s \cdot
      \epsilon}.
  \]
\end{LEMMA}
\begin{proof}
  The width of the oracle is clear: when returning a point $x = \frac{\OPT}{c_i}
  e_i$,
  \[
    A_i \cdot x - 1 \leq \frac{\OPT}{c_{min}} - 1.
  \]
  For accuracy, note that the quality score $Q$ has sensitivity at most
  \[
    \Delta = \frac{3\OPT}{c_{min} \cdot s}.
  \]
  Why? On neighboring databases, there are two possible changes: first we may
  have $| y_i -  y_i' | \leq 2/s$, and second we may have an extra term in the
  sum on one neighbor (since the sum is taken over all constraints, and one
  neighboring instance has an extra constraint). The first source
  contributes sensitivity $2 \OPT /(c_{min} s)$, and since $\|y\|_\infty,
  \|y'\|_\infty \leq 1/s$, the second source contributes  sensitivity
  $\OPT /(c_{min} s)$.

  Now, since there are $d$ possible outputs, the accuracy guarantee for the
  exponential mechanism (\Cref{exp:acc}) shows that $O$ selects a point
  with additive error at most
  \[
    \alpha = \frac{2\Delta}{\epsilon} \log d \log (1/\gamma) = \frac{6
    \OPT}{c_{min} s \epsilon} \log d \log(1/\gamma)
  \]
  with probability at least $1 - \gamma$. Hence, we are done.
\end{proof}
\fi
Now, it follows that \Cref{dual-lp-nonpriv} solves the private
fractional set cover problem with the following accuracy guarantee.
\begin{THEOREM}
  Let $\beta \in (0, 1)$. With probability at least $1 - \beta$,
  \Cref{dual-lp-nonpriv} with the exponential mechanism as an oracle
  \PREF{sc-oracle}---where $\rho$ is the width of the oracle and $\alpha \leq
  9 \rho$---finds a point $x^*$ such that $A_i x^* \geq 1 - \alpha$ except for
  at most $s$ constraints $i$, where
  \[
    s = \tilde{O} \left( \frac{\OPT^2 \log d \log^{1/2} m \log (1/\beta) \log^{1/2}
        (1/\delta)}{c^2 \cdot \alpha^2 \cdot \epsilon} \right).
  \]
  \Cref{dual-lp-nonpriv} is also $\epsilon$-high sensitivity constraint private.
\end{THEOREM}
\ifshort
\else
\begin{proof}
  Let $\epsilon'$ be as in~\Cref{cp-priv}, and let $\gamma = \beta/T$ with $T$
  as in \Cref{dual-lp-nonpriv}.  Unfolding the definition of $\epsilon'$ and
  $\rho$ and applying \Cref{sc-oracle}, the oracle gives accuracy
  \[
    \frac{6 \OPT}{c_{min} s \epsilon'} \log d \log(1/\gamma)
    = \frac{96\sqrt{2} \OPT^2 \log d \log^{1/2} m \log (1/\gamma) \log^{1/2}
      (1/\delta)}{c_{min}^2 \epsilon s \alpha}
  \]
  with probability at least $1 - \gamma$. Set this equal to $\alpha/3$. By
  assumption $\alpha \leq 9 \rho$, so \Cref{dual-lp} applies: with
  probability at least $1 - \beta$, there is a set $S$ of at most $s$
  constraints such $A_i x^* \geq 1 - \alpha$ for every $i \notin S$, where
  \[
    s = O \left( \frac{\OPT^2 \log d \log^{1/2} m \log (1/\gamma) \log^{1/2}
        (1/\delta)}{c^2 \cdot \alpha^2 \cdot \epsilon} \right).
  \]
  and $\gamma = \beta/T$.
\end{proof}
\fi
%
\ifappendix
\else
\begin{REMARK}
  A variant of the efficient private set cover problem has been investigated by
  \citet{GLMRT10}. Our techniques are more general, but the solution we provide
  here has an imcomparable accuracy guarantee. We include this example to
  demonstrate how to use \Cref{dual-lp-nonpriv} and \Cref{dual-lp}.
  \ifshort
  \else
  On the one hand, we may fail to satisfy some of the coverage constraints, and
  if we imagine that each uncovered element can be covered at a cost of $1$, our
  approximation guarantee now depends on $\OPT$ unlike the guarantee of
  \citet{GLMRT10}.

  On the other hand, we output an explicit solution whereas the algorithm of
  \citet{GLMRT10} outputs an implicit solution, a ``set of instructions'' that
  describes a set cover when paired with the private data. (Their approach can
  also be interpreted as satisfying the weaker guarantee of \emph{joint
    differential privacy} \citep{KPRU14} rather than standard differential
  privacy.) Finally, our techniques apply to general constraint-private linear
  programs, not just set cover.
  \fi
\end{REMARK}
\fi



\SECTION{Low-Sensitivity LPs}
\ifappendix
\else
Let us now turn to low-sensitivity LPs. Recall that for these LPs, the distance
between adjacent inputs decreases as the size of the database (i.e., the number
of individuals) grows. First, a few simplifying assumptions. Like above, we will
continue to solve feasibility LPs of the following form:
\begin{align*}
  &\text{find } x \in \RR^d_+ \\
  \text{s.t. } &Ax \leq b
\end{align*}
Unlike the case for general constraint private LPs, we require that the feasible
solution is a distribution, i.e., is non-negative and has $\ell_1$ norm $1$.
Note that if the optimal solution has $\ell_1$ norm $L$, then the rescaled LP
\begin{align*}
  &\text{find } x \in \RR^d_+ \\
  \text{s.t. } &Ax \leq b/L
\end{align*}
has a distribution as a solution. Our algorithms will find a point $x^*$ such
that $Ax^* \leq b/L + \alpha \cdot \mathbf{1}$, so if we set $\alpha =
\alpha'/L$, then $A(Lx^*) \leq b + \alpha'$ gives an approximate solution to the
original, unscaled LP.
\fi

\ifshort
\else
\SUBSECTION{Solving LPs with Multiplicative Weights}

Before getting into specific kinds of low-sensitivity LPs, we first review
another standard method for solving LPs via the standard multiplicative weights
algorithm presented in~\Cref{alg:MW}.
\begin{algorithm}[ht!]
  \begin{algorithmic}
    \STATE{Let $\tilde{A}^1$ be the uniform distribution on $\cA$}
    \STATE{For $t = 1,2,\dots,T$:}
    \INDSTATE[1]{Receive loss vector $\ell^t$ (may depend on $A^{1}, \dots,
    A^{t}$)}
    \INDSTATE[1]{{\bf For each} $a \in \cA${\bf :}}
    \INDSTATE[2]{Update $A^{t+1}_a = e^{- \eta \ell^t_a} \tilde{A}^{t}_a$ for every
  $a \in \cA$}
  \INDSTATE[1]{Normalize $\tilde{A}^{t+1} = A^{t + 1} / |A_{t + 1}|$}
  \end{algorithmic}
  \caption{The Multiplicative Weights Algorithm, $MW_{\eta}$}
  \LABEL{alg:MW}
\end{algorithm}

Unlike the dense multiplicative weights approach presented earlier
\PREF{dual-lp-nonpriv}, we use multiplicative weights to maintain a distribution
over the {\em variables} rather than the constraints.%
\footnote{%
  Readers familiar with game theory may notice that we are solving LPs by
  finding the equilibrium of a two player, zero-sum game. Then,
  \Cref{dual-lp-nonpriv} is solving the game with MW over the constraints and
  best response over the variables, while the approach we present in this
  section swaps the two roles. }{}
This distribution will be the candidate solution, and we define
losses by the maximum constraint violation of this candidate solution at each
step. It will be useful to first define an oracle for linear maximizations.

\begin{DEFINITION}
  For $\epsilon > 0, \gamma > 0$ an {\em $(\alpha,
  \gamma)$-dual oracle}, given $A, b, x$ as input, finds a
  constraint $i \in [m]$ such that
  \[
    A_i x - b_i \geq \max_j A_j x - b_j - \alpha,
  \]
  with probability at least $1 - \gamma$.
\end{DEFINITION}

We now give the full algorithm in~\Cref{primal-lp-nonpriv}.

\ifshort
\else
\begin{algorithm}[ht!]
  \begin{algorithmic}
    \STATE{Input $A \in \RR^{m \times d}$, $b \in \RR^m$.}
    \STATE{Let $\tilde{x}^{1}$ be the uniform distribution in $\RR^d$, $\rho =
      \max_{ij} |A_{ij}|$ be the {\em width} of the LP, $\alpha > 0$ be the
      desired accuracy. Let $\mathit{Oracle}$ be a $(\alpha, \gamma)$-dual
      oracle, and set
      \begin{mathpar}
        \eta = \sqrt{\frac{\log d}{T}}, \and
        T = \frac{9 \rho^2 \log d}{\alpha^2}.
      \end{mathpar}
    }
    \STATE{For $t = 1,\dots,T$:}
    \INDSTATE[1]{Find $p^t = \mathit{Oracle}(A, b, \tilde{x}^t)$}
    \INDSTATE[1]{Compute losses $\ell^t_i := (1/\rho) A_{p^t i}$}
    \INDSTATE[1]{Update $\tilde{x}^{t+1}$ from $\tilde{x}^t$ and $\ell^t$ via
      multiplicative weights.}
    \STATE{Output $\bar{x} = (1/T) \sum_{t = 1}^T \tilde{x}^t$}
  \end{algorithmic}
  \caption{Solving for LP feasibility with primal multiplicative weights}
  \LABEL{primal-lp-nonpriv}
\end{algorithm}
\fi

Then, the following accuracy guarantee is known.

\begin{THEOREM}[\citet{PST95}] \LABEL{primal-lp}
  Suppose there is a feasible distribution solution of the linear program $Ax
  \leq b$.  Then, running \Cref{primal-lp-nonpriv} with an $(\alpha/3,
  \gamma)$-dual oracle finds a point $x$ such that $Ax \leq b +
  \alpha \cdot \mathbf{1}$ with probability at least $1 - T\gamma$.
\end{THEOREM}
\fi

\SUBSECTION{Scalar-Private LPs} \LABEL{sec:qr}
\ifshort
We defer our scalar-private results to \APPREF{sec:qr}.
\else
First, we consider linear programs where the objective and constraint
coefficients are public data, but the right hand side in the constraints may
contain private data. Roughly, a private database $D$ maps to an objective
vector $c(D)$, a constraint matrix $A(D)$, and a vector $b(D)$. For every pair
of neighboring databases $D, D'$, we have $c(D) = c(D')$ and $A(D) = A(D')$
independent of the data, and $\|b(D) - b(D')\|_\infty \leq \Delta_\infty$. We
will think of $\Delta_\infty$ as decreasing in $n$; our accuracy guarantees will
be trivial if this is not true. As usual, we will assume the LP is in
feasibility form, and leave the objective $c$ implicit. Formally:

\begin{DEFINITION}
  A randomized algorithm $\cM$ with inputs vector $b \in \RR^m$ and matrix $A
  \in \RR^{m \times d}$, and outputting a vector in $\RR^d$ is {\em $(\epsilon,
    \delta)$-low sensitivity scalar private with sensitivity $\Delta_\infty$} if
  for any $b, b'$ such that $\|b - b'\|_\infty \leq \Delta_\infty$,
  \[
    \Pr[ \cM(b, A) \in S ] \leq e^\epsilon \Pr[ \cM(b', A') \in S] + \delta 
  \]
  for any set $S \subseteq \RR^d$.
\end{DEFINITION}

The algorithm we use is a slight generalization of the offline
private multiplicative weights algorithm \citep{GHRU11,HLM12} (building on the
influential work of \citet{HR10}, who introduced the ``online'' variant).  In
our framework, we will express the algorithm as a differentially private variant
of \Cref{primal-lp-nonpriv} to solve these linear programs while preserving
differential privacy.

Throughout, we assume that the vector $b$ is private data. On neighboring
databases, $b$ can change by at most $\Delta_\infty$ in $\ell_\infty$ norm.
Looking at \Cref{primal-lp-nonpriv}, we see that the only place we touch the
private data is in the dual oracle. Accordingly, if the dual oracle is private
in $b$, then the whole algorithm is private.


\begin{THEOREM} \LABEL{qp-priv}
  Let $\epsilon, \delta, T$ be as in \Cref{primal-lp-nonpriv}, and let
  \[
    \epsilon' = \frac{\epsilon}{\sqrt{8T \log(1/\delta)}}.
  \]
  \Cref{primal-lp-nonpriv}, run with an $\epsilon'$-private dual oracle is
  $(\epsilon, \delta)$-differentially private.
\end{THEOREM}
\begin{proof}
  Direct from composition (\Cref{composition}).
\end{proof}

Just like in private multiplicative weights for private query release, the
exponential mechanism gives an appropriate dual oracle.

\begin{LEMMA} \LABEL{qp-dual}
  Let $\epsilon, \gamma > 0$ be given, and suppose the vector $b$ can differ by
  at most $\Delta_\infty$ in $\ell_\infty$ norm on neighboring instances. Then,
  the $\epsilon$-private exponential mechanism with quality score
  \[
    Q(i, b) = A_i x - b_i
  \]
  is an $(\alpha, \gamma)$-dual oracle, for
  \[
    \alpha = \frac{2 \Delta_\infty}{\epsilon} \cdot \log \left( \frac{m}{\gamma}
    \right).
  \]
\end{LEMMA}
\begin{proof}
  This is $\epsilon$-private by definition, and the accuracy follows from the
  accuracy of the exponential mechanism (\Cref{exp:acc})---the quality score is
  $\Delta_\infty$-sensitive in $b$, and the output ranges over the constraints,
  so has size $m$.
\end{proof}

Combining the MW with the oracle, our private low-sensitivity scalar-private LP
solver~\Cref{primal-lp-nonpriv} satisfies the following accuracy guarantee.

\begin{THEOREM}
  Let $\alpha, \beta \in (0, 1)$ be given. Suppose the linear program $Ax \leq
  b$ has a distribution as a feasible solution. \Cref{primal-lp-nonpriv}, run
  with the exponential mechanism as a dual oracle \PREF{qp-dual}, is
  $(\epsilon, \delta)$-low sensitiivty scalar private with sensitivity
  $\Delta_\infty$, and finds $x^*$ satisfying $Ax^* \leq b + \alpha \cdot
  \mathbf{1}$, with probability at least $1 - \beta$, where
  \[
    \alpha = \tilde{O} \left( \frac{ \rho^{1/2} \Delta_\infty^{1/2}
      }{\epsilon^{1/2}} \cdot \log^{1/4} d \log^{1/4} (1/\delta) \log^{1/2}
        (1/\beta) \log^{1/2} m \right).
  \]
\end{THEOREM}
\begin{proof}
  Let $\epsilon'$ be as in \Cref{qp-priv}, and let $\gamma = \beta / T$ with $T$
  from \Cref{primal-lp-nonpriv}. By \Cref{qp-dual}, the $\epsilon'$-private
  exponential mechanism with quality score
  \[
    Q(i, b) = A_i x - b_i
  \]
  is an $(\alpha/3, \gamma)$-dual oracle for
  \[
    \alpha = \frac{6 \Delta_\infty \sqrt{8 T \log(1/\delta)}}{\epsilon} \cdot
    \log \left( \frac{m T}{\beta} \right)
    = \frac{18 \rho \Delta_\infty \sqrt{8 \log d \log(1/\delta)}}{\alpha \epsilon} \cdot \log
    \left( \frac{9 \rho^2 (\log d) m}{\alpha^2 \beta} \right).
  \]
  Solving,
  \[
    \alpha = \tilde{O} \left( \frac{ \rho^{1/2} \Delta_\infty^{1/2}
      }{\epsilon^{1/2}} \cdot \log^{1/4} d \log^{1/4} (1/\delta) \log^{1/2}
      (1/\beta) \log^{1/2} m \right)
\]
as desired.
\end{proof}

\begin{REMARK}
This bound generalizes the guarantee for the private multiplicative weights
algorithm when privately generating synthetic data for linear
queries~\citep{HR10}. In that setting, there is one variable for each element in
some underlying data universe $\mathcal{X}$ (and so $d = |\mathcal{X}|$), and
there is one equality constraint for each of $k$ linear queries (and so $m =
k$).
\end{REMARK}

Now, let us consider the low-sensitivity version of constraint privacy:
neighboring instances have constraint matrices that differ to a small degree.
We distinguish two further subcases: either every coefficient in each constraint
can differ, or only a few coefficients in each constraint can differ.
\fi
\SUBSECTION{Row/Matrix-Private LPs} \LABEL{sec:mp}
\ifappendix
\else
Suppose we have the feasibility problem
\begin{align*}
  &\text{find } x \\
  \text{s.t. } &Ax \leq b,
\end{align*}
where some entries in $A$ may change by at most $\Delta_\infty$ on a neighboring
instance.
\fi

\ifshort
For row privacy, we want to find a solution that hides a single constraint's low
sensitivity change.
\else
Roughly, a private database $D$ maps to an objective vector $c(D)$, a
constraint matrix $A(D)$, and a vector $b(D)$. For every pair of neighboring
databases $D, D'$, we have $c(D) = c(D')$ and $b(D) = b(D')$ independent of the
data, and $\|A(D) - A(D')\|_\infty \leq \Delta_\infty$.  Again, we will think of
$\Delta_\infty$ as decreasing in $n$; our accuracy guarantees will be trivial if
this is not true.  Our techniques work equally well whether only a single row of
$A$ or the entire matrix $A$ can differ, so we will assume the latter. We will
also assume that the LP is in feasibility form, and leave the objective $c$
implicit.
\fi%
Formally:

\begin{DEFINITION}
  A randomized algorithm $\cM$ with inputs vector $b \in \RR^m$ and matrix $A
  \in \RR^{m \times d}$, and outputting a vector in $\RR^d$ is {\em $(\epsilon,
    \delta)$-low sensitivity row private with sensitivity $\Delta_\infty$} if
  for any $A, A'$ such that $\|A - A'\|_\infty \leq \Delta_\infty$,
  \[
    \Pr[ \cM(b, A) \in S ] \leq e^\epsilon \Pr[ \cM(b', A') \in S] + \delta 
  \]
  for any set $S \subseteq \RR^d$.
\end{DEFINITION}

\ifshort
\else
We will normalize the problem so each entry in $A$ is in $[-1, 1]$.

The basic idea is to use multiplicative weights over the primal variables, with
a dual oracle selecting the most violated constraint---since the
losses fed into the multiplicative weights algorithm now depend on private data
(the matrix $A$), we add Laplace noise to the loss vectors as they are selected.
The full algorithm is given in \Cref{primal-lp-rowsens}.

\begin{algorithm}[ht!]
  \begin{algorithmic}
    \STATE{Input $A \in [-1, 1]^{m \times d}, b \in \RR^m$.}
    \STATE{Let $\tilde{x}^{1}$ be the uniform distribution in $\RR^d$, $\alpha >
    0$ be the desired accuracy, and $\Delta_\infty$ be the sensitivity. Let
    $\mathit{Oracle}$ be a $(\alpha, \gamma)$-dual oracle, and set
      \begin{mathpar}
        T = \frac{144 \log d}{\alpha^2}, \and
        \epsilon' = \frac{\epsilon}{4\sqrt{dT \log(1/\delta)}}, \and
        \eta = \sqrt{\frac{\log d }{T}}.
      \end{mathpar}
    }
    \STATE{For $t = 1,\dots,T$:}
    \INDSTATE[1]{Find $p^t = \mathit{Oracle}(A, b, \tilde{x}^t)$.}
    \INDSTATE[1]{Compute private losses $\hat{\ell}^t_i := \frac{ A_{p^t i} +
        \Lap \left( \frac{\Delta_\infty}{\epsilon'}\right)}{2}$.}
    \INDSTATE[1]{For each $i$, update $x^{t+1}_i = e^{- \eta \hat{\ell}^t_i}
      \cdot \tilde{x}_i^t$.}
    \INDSTATE[1]{Normalize $\tilde{x}^{t + 1} = x^{t + 1} / |x^{t + 1}| $.}
    \STATE{Output $\bar{x} = (1/T) \sum_{t = 1}^T \tilde{x}^t$.}
  \end{algorithmic}
  \caption{Row/matrix private LP solver}
  \LABEL{primal-lp-rowsens}
\end{algorithm}

We can now show privacy and accuracy for \Cref{primal-lp-rowsens}.

\begin{THEOREM} \LABEL{rowsens-priv}
  Let $\epsilon, \epsilon', \delta, \Delta_\infty$ be as in
  \Cref{primal-lp-rowsens}.  \Cref{primal-lp-rowsens} run with an
  $\epsilon'$-private dual oracle is $(\epsilon, \delta)$-low sensitivity row
  private with sensitivity $\Delta_\infty$.
\end{THEOREM}
\begin{proof}
  \Cref{primal-lp-rowsens} performs $dT$ Laplace operations, and $T$ oracle
  operations. Each operation is $\epsilon'$-private, so this is at most $2dT$
  $\epsilon'$-private operations. By our choice of $\epsilon'$ and
  \Cref{composition}, the whole algorithm is $(\epsilon, \delta)$-differentially
  private.
\end{proof}

We can show that the exponential mechanism is a private dual oracle.
\begin{LEMMA} \LABEL{rp-dual}
  Let $\epsilon, \gamma > 0$ be given, and suppose the matrix $A$ can differ by
  at most $\Delta_\infty$ in $\ell_\infty$ norm on neighboring instances. Let
  $x$ be any distribution. Then, the $\epsilon$-private exponential mechanism
  with quality score
  \[
    Q(i, b) = A_i x - b_i
  \]
  is an $(\alpha, \gamma)$-dual oracle, for
  \[
    \alpha = \frac{2 \Delta_\infty}{\epsilon} \cdot \log \left( \frac{m}{\gamma}
    \right).
  \]
\end{LEMMA}
\begin{proof}
  Since $x$ is a distribution, the quality score is $\Delta_\infty$-sensitive
  and accuracy follows from accuracy of the exponential mechanism
  (\Cref{exp:acc}).
\end{proof}

While previously our accuracy theorems followed from standard accuracy results
for solving LPs using multiplicative weights, the proof for
\Cref{primal-lp-rowsens} does not. Since the constraint matrix is private,
\Cref{primal-lp-nonpriv} perturbs the losses and requires a more custom
analysis. So, we will need a standard regret bound for multiplicative weights.

\begin{THEOREM}[\citet{LW94}]
  \LABEL{MW-regret}
  Let $\set{\tilde{A}^t}$ be the distributions obtained by $MW_{\eta}$ with
  arbitrary losses $\set{\ell^t}$ satisfying $\|\ell^t\|_\infty \leq 1$. Suppose
  that $\eta \leq 1/2$. Let $A^* = \mathbf{1}_{a = a^*}$, for some $a^* \in
  \cA$.  Then,
  \begin{align*}
    \sum_{t = 1}^T \< \ell^t, \tilde{A}^t \>
    &\leq
    \sum_{t = 1}^T \< \ell_t, A^* \>  + \eta + \frac{\log |\cA|}{\eta T}.
  \end{align*}
\end{THEOREM}

\fi
\ifshort
We give a solver in \APPREF{sec:mp}, with the following guarantee.
\begin{THEOREM}
  Let $\epsilon, \delta, \beta > 0$, and let $\Delta_\infty$ be the sensitivity
  of the LP. Suppose the program has a distribution as a feasible solution.
  There is an $(\epsilon, \delta)$-low sensitivity row private algorithm with
  sensitivity $\Delta_\infty$ that with probability at least $1 - \beta$,
  produces a point  $x^*$ with $Ax^* \leq b + \alpha \cdot \mathbf{1}$, where
  \[
    \alpha = \tilde{O} \left( \frac{\Delta_\infty^{1/2} d^{1/4}
      }{\epsilon^{1/2}} \cdot \polylog \left( d, m, \frac{1}{\beta},
        \frac{1}{\delta} \right) \right).
  \]
\end{THEOREM}
\else
Then, we have the following accuracy guarantee.
\begin{THEOREM} \LABEL{rowsens-acc}
  Let $\beta > 0$. Suppose the program has a distribution as a feasible
  solution. Then, with probability at least $1 - \beta$,
  \Cref{primal-lp-rowsens} run with the exponential mechanism as a dual oracle
  \PREF{rp-dual} finds a solution $x^*$ such that $Ax^* \leq
  b + \alpha \cdot \mathbf{1}$, where
  \[
    \alpha = \tilde{O} \left( \frac{\Delta_\infty^{1/2} d^{1/4}
      }{\epsilon^{1/2}} \cdot \polylog \left( d, m, \frac{1}{\beta},
        \frac{1}{\delta} \right) \right).
  \]
\end{THEOREM}
\begin{proof}
  Let $\epsilon'$ be as in \Cref{rowsens-priv}, $T$ be from
  \Cref{primal-lp-rowsens}, and $\gamma = \beta/2dT$. By \Cref{rp-dual}, with
  probability at least $1/\gamma$, the oracle's choices satisfy
  \[
    \left( A_i \cdot \tilde{x}^t - b_i\right)  - \left(
      A_{p^t} \cdot \tilde{x}^t - b_{p^t} \right) \leq \frac{2
      \Delta_\infty}{\epsilon'} \cdot \log \left( \frac{m}{\gamma} \right)
  \]
  for all constraints $i$. Note that the left hand side is equal to $\left( A_i
    \cdot \tilde{x}^t - b_i \right) - \left( \ell^t \cdot \tilde{x}^t - b_{p^t}
  \right)$. Taking a union bound over all $T$ steps, this is true for all $p^t$
  with probability at least $1 - \beta/2d \geq 1 - \beta/2$; condition on this event.

  By a tail bound on the Laplace mechanism (\Cref{lap:tail}), with probability
  at least $1 - \gamma$, a noisy loss $\hat{\ell}^t_i$ satisfies
  \[
    \left| \hat{\ell}^t_i - (1/2) \ell^t_i \right| \leq
    \frac{\Delta_\infty}{\epsilon'} \cdot \log \left( \frac{1}{\gamma}
    \right).
  \]
  Taking a union bound over all $T$ steps and $d$ losses, this is true for all
  losses with probability at least $1 - \beta/2$; condition on this event.

  We first show that if these errors are small, then the theorem holds.
  Assume
  \begin{equation} \LABEL{small-noise}
    \frac{2 \Delta_\infty}{\epsilon'} \cdot \log \left( \frac{m}{\gamma}
    \right) \leq \frac{\alpha}{6},
  \end{equation}
  so that both right hand sides above are at most $\alpha/6$. Since $\alpha <
  1$, this implies that every Laplace noise is at most $1$, so the noisy losses
  are bounded: $|\hat{\ell}^t_i| \leq 1$.

  Let $x^*$ be an exactly feasible point, with $\ell_1$ norm $1$. By the regret
  guarantee for multiplicative weights (\Cref{MW-regret}),
  \begin{align*}
    \frac{1}{T} \sum_t \hat{\ell}^t \cdot \tilde{x}^t &\leq \frac{1}{T} \sum_t
    \hat{\ell}^t \cdot x^* + \eta + \frac{\log d}{\eta T} \\
    \frac{1}{T} \sum_t \left( \frac{ A_{p^t} }{2} + \frac{\nu }{2} \right)
    \cdot \tilde{x}^t - \frac{b_{p^t}}{2}
    = \frac{1}{T} \sum_t \hat{\ell}^t \cdot \tilde{x}^t - \frac{b_{p^t}}{2}
    &\leq \frac{1}{T} \sum_t \hat{\ell}^t \cdot x^* - \frac{b_{p^t}}{2} + \eta +
    \frac{\log d}{\eta T},
  \end{align*}
  where $\nu$ is a vector of independent draws from
  $\Lap(\Delta_\infty/\epsilon')$.  Let $i$ be any constraint. Since we assumed
  the error of the exponential mechanism to be small (\Cref{small-noise}) and
  $\tilde{x}^t$ is a distribution,
  \[
    \frac{1}{T} \sum_t \left( \frac{1}{2} A_i \cdot \tilde{x}^t - \frac{b_i}{2}
    \right) + \frac{1}{2} \nu\leq \frac{1}{T} \sum_t \hat{\ell}^t \cdot x^* -
    \frac{b_{p^t}}{2} + \eta + \frac{\log d}{\eta T} + \frac{\alpha}{6}.
  \]
  By assumption (\Cref{small-noise}) $|\nu| \leq \alpha/6$, so
  \[
    \frac{1}{T} \sum_t A_i \cdot \tilde{x}^t - b_i \leq \frac{1}{T} \sum_t
    2\hat{\ell}^t \cdot x^* - b_{p^t} + \alpha/3 + \alpha/6 + \eta + \frac{\log
      d}{\eta T}.
  \]
  Since $x^*$ is a feasible point, $A_i \cdot x^* - b_i \leq 0$ for all $i$.  By
  assumption (\Cref{small-noise}), $\left| A_i - 2 \hat{\ell}^t_i  \right| \leq
  \alpha/3$. By our choice of $\eta$ and $T$,
  \begin{align*}
    A_i \cdot \bar{x} - b_i &\leq \frac{1}{T} \sum_t A_i \cdot x^* - b_i +
    \alpha/3 + \alpha/3 + \alpha/6 + \eta + \frac{\log d}{\eta T} \\
    &\leq \frac{5\alpha}{6} + \eta + \frac{\log d }{\eta T} \leq \alpha,
  \end{align*}
  as desired.

  Now, it only remains to show \Cref{small-noise}. By unfolding definitions like
  before, it suffices to take
  \[
    \alpha \geq \frac{12 \Delta_\infty^{1/2} d^{1/4} (\log d)^{1/4} (\log
      (1/\delta))^{1/4}}{\epsilon^{1/2}} \cdot \left( \log \frac{288 d \log d
        m}{\alpha^2 \beta} \right)^{1/2}.
  \]
\end{proof}
\fi

\SUBSECTION{Column Private LPs}\LABEL{sec:colp}
\jh{Is this worth talking about? It's literally the exact same algorithm with a
  $\Delta_\infty$ changed to a $\Delta_1$, giving a slightly worse dependence on
  $d$.}
\ifshort
We defer our column private results, which are quite similar to our row private
results, to \APPREF{sec:colp}.
\else
Rather than an entire row changing, neighboring LPs may differ in a {\em
  column}; that is, they differ in coefficients corresponding to a single
variable.  Roughly, a private database $D$ maps to an objective vector
$c(D)$, a constraint matrix $A(D)$, and a vector $b(D)$. For every pair of
neighboring databases $D, D'$, we have $c(D) = c(D')$ and $b(D) = b(D')$
independent of the data, and $\|A(D)_i - A(D')_i\|_\infty \leq \Delta_1$ for
every row $i$ of the constraint matrix.  Again, we will think of $\Delta_1$ as
decreasing in $n$; our accuracy guarantees will be trivial if this is not true.
Formally:

\begin{DEFINITION}
  A randomized algorithm $\cM$ with inputs vector $b \in \RR^m$ and matrix $A
  \in \RR^{m \times d}$, and outputting a vector in $\RR^d$ is {\em $(\epsilon,
    \delta)$-low sensitivity column private with sensitivity $\Delta_1$} if for
  any $A, A'$ such that $\|A_i - A_i'\|_\infty \leq \Delta_1$ for each row $i
  \in [m]$,
  \[
    \Pr[ \cM(b, A) \in S ] \leq e^\epsilon \Pr[ \cM(b', A') \in S] + \delta 
  \]
  for any set $S \subseteq \RR^d$.
\end{DEFINITION}

We can use a very slight modification of \Cref{primal-lp-rowsens} to solve these
LPs privately; the algorithm is given in \Cref{primal-lp-colsens}.

\begin{algorithm}[ht!]
  \begin{algorithmic}
    \STATE{Input $A \in [-1, 1]^{m \times d}, b \in \RR^m$.}
    \STATE{Let $\tilde{x}^{1}$ be the uniform distribution in $\RR^d$, $\alpha >
    0$ be the desired accuracy, and $\Delta_1$ be the sensitivity. Let
    $\mathit{Oracle}$ be an $(\alpha, \gamma)$-dual oracle, and set
      \begin{mathpar}
        T = \frac{144 \log d}{\alpha^2}, \and
        \epsilon' = \frac{\epsilon}{4\sqrt{T \log(1/\delta)}}, \and
        \eta = \sqrt{\frac{\log d }{T}}.
      \end{mathpar}
    }
    \STATE{For $t = 1,\dots,T$:}
    \INDSTATE[1]{Find $p^t = \mathit{Oracle}(A, b, \tilde{x}^t)$.}
    \INDSTATE[1]{Compute private losses $\hat{\ell}^t_i := \frac{A_{p^t i} +
        \Lap \left( \frac{\Delta_1}{\epsilon'}\right)}{2}$.}
    \INDSTATE[1]{For each $i$, update $x^{t+1}_i = e^{- \eta \hat{\ell}^t_i}
      \cdot \tilde{x}_i^t$.}
    \INDSTATE[1]{Normalize $\tilde{x}^{t + 1} = x^{t + 1} / |x^{t + 1}| $.}
    \STATE{Output $\bar{x} = (1/T) \sum_{t = 1}^T \tilde{x}^t$.}
  \end{algorithmic}
  \caption{Column private LP solver}
  \LABEL{primal-lp-colsens}
\end{algorithm}

Like before, we can show that the exponential mechanism can be used as a
private dual oracle.

\begin{LEMMA} \LABEL{cp-dual}
  Let $\epsilon, \gamma > 0$ be given, and suppose neighboring matrices $A, A'$
  satisfy $\| A_i - A_i' \|_1 \leq \Delta_1$ for every row $i$. Let $x$ be any
  distribution. Then, the $\epsilon$-private exponential mechanism with quality
  score
  \[
    Q(i, b) = A_i x - b_i
  \]
  is an $(\alpha, \gamma)$-dual oracle, for
  \[
    \alpha = \frac{2 \Delta_1}{\epsilon} \cdot \log \left( \frac{m}{\gamma}
    \right).
  \]
\end{LEMMA}
\begin{proof}
  Since $x$ is a distribution, the quality score is $\Delta_1$-sensitive
  and accuracy follows from accuracy of the exponential mechanism
  (\Cref{exp:acc}).
\end{proof}

\begin{THEOREM} \LABEL{colsens-priv}
  Let $\epsilon, \epsilon', \delta, \Delta_1$ be as in \Cref{primal-lp-colsens}.
  \Cref{primal-lp-colsens} run with an $\epsilon'$-private dual oracle is
  $(\epsilon, \delta)$-low sensitiivty column private with sensitivity
  $\Delta_1$.
\end{THEOREM}
\begin{proof}
  Since the loss vector $\ell^t$ can differ by at most $\Delta_1$ in $\ell_1$
  norm, adding Laplace noise with scale $\Delta_1/\epsilon$ suffices for
  $\epsilon$-differentially privacy. Thus, there are $T$ Laplace and oracle
  mechanism steps, each $\epsilon'$-private. By choice of $\epsilon'$ and
  composition, \Cref{primal-lp-colsens} is $(\epsilon, \delta)$-differentially
  private.
\end{proof}

\begin{THEOREM} \LABEL{colsens-acc}
  Let $\beta > 0$. Suppose the program has a distribution as a feasible
  solution. Then, with probability at least $1 - \beta$,
  \Cref{primal-lp-colsens} run with the exponential mechanism \PREF{cp-dual}
  as oracle finds a point $x^*$ such that $Ax^* \leq b + \alpha \cdot
  \mathbf{1}$, where
  \[
    \alpha = \tilde{O} \left( \frac{\Delta_1^{1/2}}{\epsilon^{1/2}} \cdot
      \polylog \left( d, m, \frac{1}{\beta}, \frac{1}{\delta} \right) \right).
  \]
\end{THEOREM}
\begin{proof}
  Let $\epsilon'$ be as in \Cref{colsens-priv}, $T$ be as
  in\Cref{primal-lp-colsens}, and $\gamma = \beta/2dT$. Letting the dual oracle
  by the $\epsilon'$-private exponential mechanism, the proof is nearly
  identical to \Cref{rowsens-acc}. The main difference is that we need
  \[
    \frac{2 \Delta_1}{\epsilon'} \cdot \log \left( \frac{1}{\gamma}
    \right) \leq \frac{\alpha}{6}
  \]
  for everything to go through. By unfolding definitions, it suffices to take
  \[
    \alpha \geq \frac{12 \Delta_1^{1/2} (\log d)^{1/4} (\log
      (1/\delta))^{1/4}}{\epsilon^{1/2}} \cdot \left( \log \frac{288 m \log d
      }{\alpha^2 \beta} \right)^{1/2}.
  \]
\end{proof}
Comparing the two previous algorithms, note $\Delta_\infty \leq \Delta_1 \leq d
\Delta_\infty$. \Cref{primal-lp-rowsens} performs better when the right
inequality is tighter, i.e., when all the coefficients in a row can differ by a
small amount. In contrast, \Cref{primal-lp-colsens} performs better when the
left inequality is tighter, that is, when a few coefficients in a row can differ
by a larger amount.
\fi

\SUBSECTION{Objective Private LPs} \LABEL{sec:op}

\ifappendix
We can show accuracy and privacy for a randomized response approach to solving
objective private LPs.
\else
For our final type of low-sensitivity LP, we consider linear programs with
objectives that depend on private data. We show that a very simple
approach---{\em randomized response}---can solve these types of LPs accurately.
Throughout, we will assume that the optimal solution to the LP has $\ell_1$
weight equal to $1$. We start with an LP in general form:
\begin{align*}
  &\max c^\top x \\
  \text{s.t. } &Ax \leq b,
\end{align*}
\fi
On instances corresponding to neighboring database $D, D'$, the objective may
change by $\Delta_1$ in $\ell_1$ norm: $\|c(D) - c(D')\|_1 \leq \Delta_1$.
Formally:

\begin{DEFINITION}
  A randomized algorithm $\cM$ with inputs vectors $b \in \RR^m$, $c \in \RR^d$
  and matrix $A \in \RR^{m \times d}$, and outputting a vector in $\RR^d$ is
  {\em $(\epsilon, \delta)$-low sensitivity objective private with sensitivity
    $\Delta_1$} if for any $c, c'$ such that $\|c - c'\|_1 \leq \Delta_1$,
  \[
    \Pr[ \cM(c, b, A) \in S ] \leq e^\epsilon \Pr[ \cM(c', b, A) \in S] + \delta 
  \]
  for any set $S \subseteq \RR^d$.
\end{DEFINITION}

\ifshort
\else
For a concrete case, a single objective coefficient may change by $\Delta_1$.
All other parts of the LP do not change: $A(D) = A(D')$, and $b(D) = b(D')$. If
we add Laplace noise to the objective and solve the resulting LP, we will get an
almost optimal, exactly feasible solution.
\fi

\begin{THEOREM}
  Suppose an objective private LP has optimal objective $\OPT$, and has optimal
  solution with $\ell_1$ weight $1$. Define
  \[
    \hat{c} = c + \Lap \left( \frac{\Delta_1 \sqrt{8d \log(1/\delta)}}{\epsilon}
      \right)^d,
  \]
  where the noise is $d$ independent draws from the Laplace distribution with
  the given parameter. Then, releasing the perturbed LP
  \begin{align*}
    &\max \hat{c}^\top x \\
    \text{s.t. } &Ax \leq b \quad \text{and} \quad \mathbf{1}^\top x = 1
  \end{align*}
  is $(\epsilon, \delta)$-low sensitivity objective private with sensitivity
  $\Delta_1$. With probability $1 - \beta$, solving the perturbed LP
  non-privately yields a point $x^*$ such that $Ax^* \leq b$ and $c^\top x^*
  \geq \OPT - \alpha$, where
  \[
    \alpha = \frac{4 \Delta_1 \sqrt{8d \log(d/\delta)}}{\epsilon}.
  \]
\end{THEOREM}
\ifshort
\else
\begin{proof}
  Since the $\ell_1$ sensitivity of $c$ is $1$ and $d$ numbers are released,
  $(\epsilon, \delta)$-privacy follows from the composition theorem
  (\Cref{composition}).%
  \footnote{%
    This is similar to the case of privately releasing {\em histogram queries}.}

  For the accuracy, note that with probability at least $1 - \beta/d$, a single
  draw of the Laplace distribution is bounded by
  \[
    \frac{\alpha}{2} = \frac{2 \Delta_1 \sqrt{8d \log(d/\delta)}}{\epsilon}.
  \]
  By a union bound, this happens with probability at least $1 -
  \beta$ for all $d$ draws; condition on this event. Then, note that if $x^*$ is
  the optimal solution to the original LP, then it is also a feasible solution to
  the perturbed LP. Let $\hat{x}^*$ be the optimal solution of the perturbed LP. Since
  the noise added to each objective coefficient is bounded by $\alpha/2$, if
  \[
    c^\top \hat{x}^* < \OPT - \alpha
  \]
  then
  \[
    \hat{c}^\top \hat{x}^* < \OPT - \alpha/2
    \quad \text{but also} \quad
    \hat{c}^\top x^* \geq \OPT - \alpha/2,
  \]
  contradicting optimality of $\hat{x}^*$ in the perturbed program. Thus, this
  algorithm finds an exactly feasible, $\alpha$-optimal solution.
\end{proof}
\fi

\SECTION{Lower Bounds} \LABEL{sec:lowerbounds}

\ifshort
We defer details of our lower bounds to \APPREF{sec:lowerbounds}.
\else
Now that we have considered various low-sensitivity LPs, let us turn to
high-sensitivity LPs. In this section, we show that most high-sensitivity LPs
cannot be solved privately to non-trivial accuracy. The exception is constraint
private LPs---as we saw \PREF{sec:cp}, these can be solved in a relaxed sense.
Our lower bounds are all reductions to reconstruction attacks: as the following
theorem shows, differential privacy precludes reconstructing a non-trivial
fraction of a database. The idea of reconstruction being a key feature of
privacy violation is due to~\citet{DN03}. The following theorem is folklore; we
provide a proof for completeness.

\begin{THEOREM} \LABEL{thm:reconstruction}
  Let mechanism $\cM : \{0,1\}^n \rightarrow [0,1]^n$ be $(\epsilon,
  \delta)$-differentially private, and suppose that for all database $D$, with
  probability at least $1-\beta$, $\|\cM (D) - D \|_1 \leq \alpha n$. Then,
  \[
    \alpha \geq \frac{1}{2} - \frac{e^\epsilon + \delta}{2 (1+e^\epsilon) (1-\beta)}
    := c(\epsilon, \delta, \beta).
  \]
  The same is true even if $D$ is restricted to have exactly $n/2$ zero entries.
\end{THEOREM}
\begin{proof}
  If we have $\cM$ as in the hypothesis, then we can round each entry of
  $\cM(D)$ to $\{0, 1\}$ while preserving $(\epsilon, \delta)$-differential
  privacy. Note that by assumption $\|\cM(D) - D\|_1 \leq \alpha n$, so the
  number of entries $\cM(D)_i$ that are more than $1/2$ away from $D_i$ is at
  most $2\alpha n$. Thus, rounding reconstructs a database in $\{0, 1\}^n$ at
  most $2\alpha n$ distance from the true database in $\ell_1$ norm; hence we
  may assume that $\cM(D) \in \{0, 1\}^n$ with $\ell_1$ norm at most $2\alpha
  n$ from $D$.

  Assume $n$ is even; we prove the case where the input database $D$ has exactly
  $n/2$ zero entries. Let $D\in \{0,1\}^n$ have exactly $n/2$ zero entries, and
  sample an index $i$ such that $D_i = 1$, and an index $j$ such that $D_j = 0$,
  both uniformly at random from $[n]$.  Let $D'$ be identical to $D$ except with
  bits $i$ and $j$ swapped.  By assumption, we have that with probability at
  least $1-\beta$
  \[
    \| \cM(D) - D \|_1 \leq 2\alpha n
    \qquad \text{and} \qquad
    \| \cM(D') - D' \|_1 \leq 2\alpha n.
  \]
  Since $i$ is chosen uniformly, we also know
  \[
    \Pr[ \cM(D)_i = D_i ] \geq (1 - 2\alpha)(1 - \beta)
    \qquad \text{and} \qquad
    \Pr[ \cM(D')_i = D'_i ] \geq (1 - 2\alpha)(1 - \beta).
  \]
  Hence, $\Pr[ \cM(D')_i = D_i ]  \leq 1 - (1 - 2\alpha)(1- \beta)$ because $D_i
  \neq D_i'$.  By $(\epsilon, \delta)$-differential privacy, we get
  \[
    (1 - 2\alpha)(1 - \beta)  \leq \Pr[\cM(D)_i = D_i ] \leq
    e^\epsilon \Pr[\cM(D')_i = D_i] + \delta \leq e^\epsilon (1 - (1
    - 2\alpha)(1 - \beta)) + \delta.
  \]
  Finally,
  \[
    1 - 2\alpha \leq \frac{e^\epsilon + \delta}{(1+ e^\epsilon)(1 - \beta)},
  \]
  as desired.
\end{proof}

For each type of impossible private LP, we show how to convert a database $D \in
\{0, 1\}^n$ to a LP, such that neighboring databases $D, D'$ lead to neighboring
LPs. We then show that a LP solver that privately solves this LP to non-trivial
accuracy leads to a reconstruction attack on $D$, violating
\Cref{thm:reconstruction}.

First, some notation. For the general LP
\begin{align*}
  &\max c^\top x \\
  \text{s.t. } &Ax \leq b,
\end{align*}
we say that $x^*$ is an {\em $\alpha$-feasible} solution if $Ax^* \leq b + \alpha
\cdot \mathbf{1}$. Likewise, we say that $x^*$ is an {\em $\alpha$-optimal} solution
if it is feasible, and
\[
  c^\top x^* \geq \max_{x : Ax \leq b} c^\top x - \alpha.
\]

\SUBSECTION{High-Sensitivity Scalars} \LABEL{sec:highsens-const}
Consider a database $D \in \{0, 1\}^n$, and the following LP:
\begin{align*}
  &\text{find } x \\
  \text{s.t. } &x_i = D_i \quad \text{for each } i
\end{align*}
%

Note that changing a single bit in $D$ will change a single right hand side in
a constraint by $1$.

\begin{THEOREM} \LABEL{lb-qp}
  Suppose mechanism $\cM$  is $(\epsilon, \delta)$-high sensitivity scalar
  private, and with probability at least $1 - \beta$, finds an $\alpha$-feasible
  solution. Then, $\alpha \geq 1/2$.
\end{THEOREM}
\begin{proof}
  Consider the gadget LP above. Note that if $\cM$  guarantees $\alpha < 1/2$,
  then $|x_i - D_i| < 1/2$ so rounding $x_i$ to $0$ or $1$ will reconstruct
  $D_i$ exactly. By \Cref{thm:reconstruction}, this is impossible under
  differential privacy.
\end{proof}

\SUBSECTION{High-Sensitivity Objective} \LABEL{sec:highsens-obj}
Consider a database $D \in \{0, 1\}^n$ with exactly $n/2$ zeros, and the
following LP:
\begin{align*}
  &\text{maximize } \sum_i D_i x_i - n/2 \\
  \text{s.t. } &\sum_i x_i = n/2, \qquad x_i \in [0, 1]
\end{align*}
Note that swapping a zero and a non-zero bit in $D$ will change exactly two
objective coefficients in the LP by $1$.  Observe that this is similar to the
objective private LP \PREF{sec:op} because we are only allowing the objective
to change.  However here we consider the setting where a single objective
coefficient changes arbitrarily, rather than by a small amount.

\begin{THEOREM} \LABEL{lb-op}
  Suppose mechanism $\cM$  is $(\epsilon, \delta)$-high-sensitivity objective
  private, and with probability at least $1 - \beta$, finds an exactly feasible,
  additive $(\alpha n)$-optimal solution.  Then, $\alpha \geq c(2\epsilon,
  \delta(1 + e^\epsilon), \beta)$.
\end{THEOREM}
\begin{proof}
  Consider the gadget LP above. Note that the optimal solution is $x_i = D_i$,
  with objective $0$. With probability at least $1 - \beta$, $\cM$ finds a
  solution $x^*$ with objective at least $-\alpha n$. In this case, $x^*$ places
  at most $\alpha n$ mass on indices with $D_i = 0$, so at least $(1 - \alpha)
  n$ mass of $D$ and $x^*$ are shared. Thus,
  \[
    \| D - x^* \|_1 \leq \alpha n.
  \]
  Since a change in $D$ leads to a distance two change in the LP, the
  composition is $(2 \epsilon, \delta (1 + e^\epsilon))$-private. By
  \Cref{thm:reconstruction}, $\alpha \geq c(2\epsilon, \delta(1 + e^\epsilon),
  \beta)$.
\end{proof}

\SUBSECTION{High-Sensitivity Constraints/Columns} \LABEL{sec:highsens-constr}
Consider a database $D \in \{0, 1\}^n$ with exactly $n/2$ zeros, and the
following LP:
\begin{align*}
  &\text{find } x_i \\
  \text{s.t. } &\sum_i D_i x_i = n/2 \\
  &x_i \in [0, 1] \quad \text{and} \quad \sum_i x_i = n/2
\end{align*}
Note that changing a single bit in $D$ will change coefficients in a single
constraint in the LP by $1$. Observe that this is similar to the column private
LP \PREF{sec:colp} because we are allowing the coefficients for a single
variable to change.  However here we consider the setting where this coefficient
can change arbitrarily, rather than by only a small amount.

This problem is also a special case of constraint private LPs \PREF{sec:cp}
because the coefficients in one (i.e., the only) constraint can change
arbitrarily. In the current setting, we want a solution that approximately
satisfies all constraints, rather than just satisfying most of the constraints.

\begin{THEOREM} \LABEL{lb-cp}
  Suppose mechanism $\cM$  is $(\epsilon, \delta)$-high-sensitivity constraint
  private, and finds an $\alpha$-feasible solution that satisfies all public
  constraints with probability at least $1 - \beta$.  Then, $\alpha \geq
  c(2\epsilon, \delta(1 + e^\epsilon), \beta)$.
\end{THEOREM}
\begin{proof}
  Consider the gadget LP above. Suppose with probability $1 - \beta$, $\cM$
  finds $x^*$ such that $\| Ax^* - b\|_\infty \leq \alpha$ for the gadget LP.
  By reasoning analogous to \Cref{lb-op}, at least $(1 - \alpha) n$ of the mass
  of $x^*$ will coincide with $D$, hence $\alpha \geq c(2\epsilon, \delta(1 +
  e^\epsilon), \beta)$ by \Cref{thm:reconstruction}.
\end{proof}

Also note that the LPs produced by this reduction differ only in the
coefficients corresponding to two variables. Hence, \Cref{lb-cp} also shows that
privately solving column private LPs to non-trivial accuracy is impossible. It's
possible that a relaxed solution, similar to allowing unsatisfied constraints in
the constraint-private case, could be possible under column privacy.

However, it is not enough to allow some constraints to be unsatisfied. Since we
can simply duplicate the constraint in our lower bound gadget multiple times,
producing a solution satisfying any {\em single} constraint to non-trivial
accuracy is impossible under high-sensitivity column privacy. A different
relaxation would be needed for non-trivial accuracy under column privacy.

\fi

\ifshort
\else
\SECTION{Discussion} \LABEL{sec:conc}
In this paper, we have initiated the study of privately solving general linear
programs. We have given a taxonomy of private linear programs, classified by
how the private data affects the LP. For each type of linear program in the
taxonomy, we have either given an efficient algorithm, or an impossibility
result.

One natural question is, to what extent do our results extend to other private
convex programs, e.g., semidefinite programs (SDPs)? A tempting approach is to
to use the {\em Matrix Multiplicative Weights} algorithm of \citet{AK07} for
solving SDPs. However, certain features of the standard multiplicative weights
algorithm which we use crucially---such as compatibility with Bregman
projections---seem more delicate  when using Matrix Multiplicative Weights.
\fi

\bibliographystyle{plainnat}

\ifshort\ssmall\fi
\bibliography{./refs}
\normalsize

\ifcamera
\else
\ifshort
\shortfalse
\appendixtrue
\labeltrue
\appendix

\fi
\fi

\end{document}